\documentclass[prb,twocolumn]{revtex4-1}
\usepackage{times}
\usepackage{bbm}
\usepackage[dvips]{graphicx}
\usepackage{latexsym,amsmath,amssymb,bm,euscript}
\usepackage[dvips]{color}
\usepackage{multirow}
\usepackage{braket}

\def\zn{\mathbb{Z}_N}
\def\znn{\mathbb{Z}_N \times \mathbb{Z}_N}

\def\zrr{\mathbb{Z}_r \times \mathbb{Z}_r}
\def\GCD{\tilde{d}}
\def\zz{\xi}
\def\fl{\mathcal{L}}
\def\fr{\mathcal{R}}
\begin{document}

\title{Exact Models for Symmetry-Protected Topological Phases in One Dimension}
\date{\today}

\author{Scott D. Geraedts}
\author{Olexei I. Motrunich}
\affiliation{Department of Physics, California Institute of Technology, Pasadena, California 91125, USA}

\begin{abstract}
We present an exactly solvable model for one-dimensional symmetry-protected topological phases with $\znn$ symmetry. The model works by binding point topological defects (domain walls) of one symmetry to charges of the other and condensing these bound states. Binding single topological defects to charges leads to symmetry-protected topological phases, while binding multiple topological defects to charges leads to phases with a combination of symmetry-breaking and topological properties.  
\end{abstract}
\maketitle

\section{Introduction}
Over the past few years, Symmetry Protected Topological phases (SPTs) have generated a lot of research interest.\cite{WenScience,WenDummies,Furusaki} Much of this interest has been devoted to the subclass of SPTs which are constructed of weakly interacting fermions, in particular the topological insulators.\cite{HasanKaneRMP,QiZhangRMP} One advantage of studying such systems is that one can make progress by applying well-known techniques from band theory. When an SPT is composed of strongly interacting particles, new theoretical techniques are needed. 
One approach is to construct models of SPTs by condensing bound states of charges and point topological defects. We have used this approach successfully in the past to construct models for bosonic interacting versions of the quantum Hall effect in two dimensions\cite{FQHE} and topological insulator in three dimensions.\cite{SO34D} 

\begin{figure}[b]
\includegraphics[width=\linewidth]{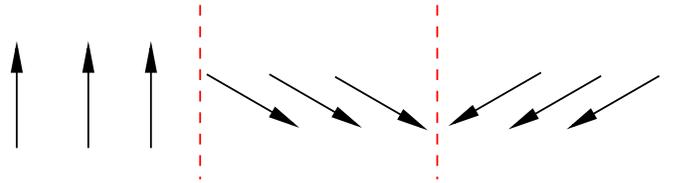}
\caption{The dashed lines are examples of domain walls in a model of $\mathbb{Z}_3$ variables.}
\label{domainwall}
\end{figure}

In this work we apply this philosophy to one-dimensional systems with discrete symmetry. The relevant point topological defect is a domain wall, pictured in Fig.~\ref{domainwall}. The system has $\zn \times \zn$ symmetry, and its ground state can be viewed as a condensate of objects which are bound states of charges of one $\zn$ symmetry and domain walls of the other. Such a model with $N=2$ was presented by Ref.~\onlinecite{DecDomWalls}. The present work extends this model to general $N$ and phases beyond purely SPT phases. 
The $\mathbb{Z}_2\times\mathbb{Z}_2$ case of our model realizes the same topological phase as the $SU(2)$ AKLT chain,\cite{AKLT1,*AKLT2} and the phase realized by the $\mathbb{Z}_3\times\mathbb{Z}_3$ version of our model has also been realized in an $SU(3)$ AKLT-like model.\cite{Furusaki} 
Models with $\znn$ symmetry can have $N$ different topological phases,\cite{WenPRB} and our model can realize all of them by binding domain walls to different numbers of charges.

It is also possible to consider binding multiple domain walls to a single charge. In higher dimension this has led to phases with intrinsic topological order, also known as Symmetry Enriched Topological (SET) phases. In this one dimensional case such topological order is not possible. Instead we find that binding $d$ domain walls (with $d$ a divisor of $N$) to each charge partially breaks the symmetry from $\znn$ to $\zrr$, where $r=N/d$. In this case $r$ different topological phases are possible,\cite{Quella}  and our model can realize all of them as well. 

\section{Model}
\begin{figure}
\includegraphics[width=\linewidth]{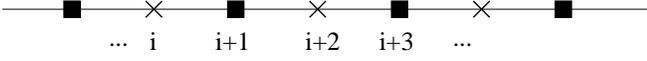}
\caption{The Hilbert space of the model. There is one species of $\mathbb{Z}_N$ variables on lattice sites with odd labels, and another species on sites with even labels.}
\label{chain}
\end{figure}

The Hilbert space of our model is shown in Figure \ref{chain}. 
We have a chain of $\zn$ variables, which can be divided into those which live on odd-indexed sites and those which live on even-indexed sites. For each variable, we write the Hamiltonian in terms of $\zn$ generalizations of the Pauli matrices, which we call $Z$ and $X$. Matrix representations of $Z$ and $X$ can be found in Ref.~\onlinecite{Furusaki}. They have the following properties:
\begin{equation}
X^{\dagger}=X^{N-1}, ~~~~ X^N=\mathbbm{1},
\label{Xprop}
\end{equation}
and similarly for $Z$. They also have the following commutation relation:
\begin{equation}
XZ=\omega ZX, ~~~\omega\equiv e^{i2\pi /N}.
\end{equation}
Our Hamiltonian is as follows:
\begin{eqnarray}
&&H=H_{\rm odd}+H_{\rm even},\nonumber\\
&&H_{\rm odd}=-\frac{1}{2}\sum_{i=2j+1}\left[ (Z_{i-1}^{\dagger})^c X_{i}^d Z_{i+1}^c + h.c.\right],\label{ham1}\\ 
&&H_{\rm even}=-\frac{1}{2}\sum_{i=2j}\left[ Z_{i-1}^c X_{i}^d (Z_{i+1}^\dagger)^c + h.c.\right].\nonumber
\end{eqnarray}
Here $c$ and $d$ are integers in $[0,1,...,N-1]$. One can check that all the terms in this Hamiltonian commute. 

The Hamiltonian also commutes with the operators
\begin{equation}
\Theta_{\rm odd}\equiv \prod_{i=2j+1} X_i, ~~~~ \Theta_{\rm even}\equiv \prod_{i=2j} X_i,
\label{symmetry}
\end{equation}
 which generate the $\znn$ symmetry.

When either $c=0$ or $d=0$ the Hamiltonian is clearly topologically trivial, but we will show that other values of $c$ and $d$ lead to topological phases. A domain wall in the order parameter of one `species' (even or odd) can be detected with operators like $Z_{i-1}^\dagger Z_{i+1}$; coupling these operators to $X_i$ operators has the effect of binding these domain walls to charged particles of the other species. This is how our Hamiltonian realizes the physical mechanism behind SPT phases discussed in the introduction. 

Let us now try to find the ground state of this Hamiltonian. Here and below we will work in the $Z$ basis, such that: 
\begin{eqnarray}
Z\ket{\zz}=\zz\ket{\zz},&&~~\zz=e^{i2\pi m/N},~~m=0,1,...,N-1,\vspace{8pt}\\
&&X\ket{\zz}=\ket{\zz \omega^{-1}}.
\end{eqnarray}
 We will first study the case with periodic boundary conditions. Consider the following wave function:
\begin{eqnarray}
&&\ket\Psi=\sum_{\{\zz_1,...,\zz_{L}\}} \alpha[\zz_1, ... , \zz_{L}]
\ket{\zz_1,..., \zz_{L}}.
\label{psigen}
\end{eqnarray}
Here $L$ is the length of the system (which is assumed to be even), and since we have periodic boundary conditions $\zz_{L+1}\equiv\zz_1$.

Consider acting with one of the terms of the Hamiltonian on the above wave function, for example a term from $H_{\rm odd}$ where the $X$ operator acts on a site index $i$. The result is:
\begin{eqnarray}
(Z_{i-1}^{\dagger})^c X_{i}^d Z_{i+1}^c \ket\Psi&=&\!\!\!\!\sum_{\{\zz_1,...,\zz_{L}\}} \!\!\!\!(\zz_{i-1}^*)^{c} (\zz_{i+1})^c 
\alpha[\zz_1,..., \zz_{L}]\nonumber\\
&&\ket{\zz_1,...,\zz_i\omega^{-d},..., \zz_{L}}.
\label{oneoperator}
\end{eqnarray}
The state $\ket{\Psi}$ is an eigenstate of this term with eigenvalue $\lambda_i$ if, for all spin configurations $\{\zz_1,...,\zz_{L}\}$:
\begin{eqnarray}
&&\lambda_i\alpha[\zz_1,...,\zz_i\omega^{-d},..., \zz_{L}]=\nonumber\\
&&=(\zz_{i-1}^*)^{c} (\zz_{i+1})^c\alpha[\zz_1,...,\zz_i, ... ,\zz_{L}].
\label{constraint}
\end{eqnarray}
A similar condition must be satisfied for $\ket{\Psi}$ to be an eigenstate of $H_{\rm odd}$, though we can see from the Hamiltonian that the $\zz_{i+1}$ variable with be complex conjugated. The periodicity of the $X_i$ operators sets some constraints on the $\lambda_i$. Applying the above operator $N$ times gives us the constraint that $\lambda_i^N=1$. 
The Hamiltonian on each site has the eigenvalue $-(\lambda_i+\lambda_i^*)/2$ [after also including the Hermitian conjugate to the operator in Eq.~(\ref{oneoperator})]. The ground state is the state with all $\lambda_i=1$, and the $\alpha$'s chosen to satisfy Eq.~(\ref{constraint}). We will discuss properties of this ground state below.

\section{SPT Phases with No Symmetry Breaking}
Models with different $c$ and $d$ can realize different phases, and both topology and symmetry breaking must be used to characterize them. We begin by considering the case where only one domain wall is bound to each charge, and there is no symmetry breaking. This happens when $d$ and $N$ are mutually prime. We can identify the topological nature of the phase by computing its projective representation of the global symmetries on a system with open boundary conditions. The method for doing this is well-known,\cite{TurnerVishwanath} but briefly summarized here. We begin by decomposing the symmetry operators in Eq.~(\ref{symmetry}) into left and right parts, $\Theta_{\rm odd}=\fl_{\rm odd}\fr_{\rm odd}$, $\Theta_{\rm even}=\fl_{\rm even}\fr_{\rm even}$, which act only at the ends of the open chain. These operators may belong to a projective representation of $\znn$. To see this we compute $\fl_{\rm even}\fl_{\rm odd}\fl^{-1}_{\rm even}\fl^{-1}_{\rm odd}\equiv \gamma$. If two phases of matter have different $\gamma$ then they are topologically distinct phases; the topologically trivial phase has $\gamma=1$. 

To put our Hamiltonian in Eq.~(\ref{ham1}) on an open chain with sites labelled from $1$ to $L$, we only allow terms which are centered on sites from $2$ to $L-1$. 
To compute the projective representation, consider the action of $\Theta_{\rm odd}$ on the ground state. We will first consider the case where $d=1$. For concreteness, we will assume that the length of the chain, $L$, is even. $\Theta_{\rm odd}$ has $X$ operators on the odd sites, and we are free to insert identities of the form $Z_{2j}^c(Z_{2j}^\dagger)^c$ on the even sites. After doing this $\Theta_{\rm odd}$ is a product of terms $(Z^\dagger_{i-1})^cX_iZ^c_{i+1}$, which in the ground state have eigenvalue $1$ and can therefore be ``removed''. On a periodic chain all of $\Theta_{\rm odd}$ can be removed in this way and therefore $\Theta_{\rm odd}\ket{\Psi_0}=\ket{\Psi_0}$; we can also argue from this that the ground state is unique. On an open chain there are no terms in the Hamiltonian which are centered at the ends of the chain, therefore after removing all terms of the form $(Z_{i-1}^\dagger)^cX_iZ_{i+1}^c$, we are still left with terms on the ends:
\begin{equation}
\Theta_{\rm odd}\ket{\Psi_0}=X_1Z_2^c(Z_L^\dagger)^c\ket{\Psi_0},
\end{equation}
and therefore
\begin{equation}
\fl_{\rm odd}=X_1Z_2^c, ~~~~~ \fr_{\rm odd}=(Z_L^\dagger)^c.
\end{equation}
Similarly
\begin{equation}
\fl_{\rm even}=(Z_1^\dagger)^c, ~~~~~ \fr_{\rm odd}=Z_{L-1}^c X_L.
\end{equation}
Using the above and Eq.~(\ref{Xprop}) we can easily see that:
\begin{equation}
\fl_{\rm even}\fl_{\rm odd}\fl_{\rm even}^{-1}\fl_{\rm odd}^{-1}=\omega^c,
\label{gamma1}
\end{equation}
and a similar result holds for the $\fr$ operators. This result implies that the ground state of the model realizes a topological phase when $c\neq 0$ and $d=1$. Furthermore, since there are $N$ different choices for $c$ the model can realize $N$ different topological phases, in agreement with the literature.\cite{WenPRB} 

The above projective represention tells us that on an open chain, there are degenerate states at each end of the chain. If $c$ and $N$ are mutually prime there are $N$ degenerate states. In general the number of degenerate states on each end is given by $N/\tilde{c}$, with $\tilde{c}$ the greatest common divisor of $c$ and $N$, $\tilde{c}\equiv \gcd(c,N)$. We can also directly see this from our Hamiltonian. On an open chain the sites $1$ and $L$ have no $X$ operator acting on them, and their values can be fixed for any eigenstate. This gives an $N$-fold degeneracy associated with each end site. However this degeneracy may not be stable to perturbations. When $\tilde{c}>1$, the following symmetry allowed perturbation is possible, and commutes with the rest of the Hamiltonian:
\begin{equation}
\delta H=-h_1X_1^{N/\tilde{c}}.
\end{equation}
This perturbation reduces the degeneracy to $N/\tilde{c}$ for each edge, in agreement with the robust prediction from Eq.~(\ref{gamma1}).

When $d>1$ and mutually prime with $N$, the same $N$ topological phases are realized. To see this, we again start with $\Theta_{\rm odd}$ but insert identities on the even sites of the form $Z_{2j}^{cs}(Z^\dagger_{2j})^{cs}$, where $s$ is an integer such that $sd=1$ (all arithmetic is done mod $N$). If $N$ and $d$ are mutually prime then $s$ exists and is mutually prime with $N$. Since $X_i^{ds}=X_i$, $\Theta_{\rm odd}$ becomes a product of terms from the Hamiltonian, raised to the $s$-th power, and such terms still have eigenvalue $1$. Similar arguments to those above then give:
\begin{equation}
\fl_{\rm odd}=X_1Z_2^{cs}, ~~ \fl_{\rm even}=(Z_1^\dagger)^{cs}, ~~ \gamma=\omega^{cs},
\end{equation}
for $\Theta_{\rm odd}$ acting on the ground state. Since $s$ is mutually prime with $N$, the $N$ different possible $c$ generate the same $N$ topological phases as the $d=1$ case.

\section{Phases with both Symmetry Breaking and SPT Order}
Finally we must deal with the case where multiple domain walls bind to each charge. In this case the resulting phases have both symmetry-breaking and topological content. Multiple domain walls bind to a charge whenever $d$ and $N$ are not mutually prime, i.e. whenever 
\begin{equation}
\tilde{d}\equiv\gcd(N,d)>1.
\end{equation}
In this case, if we start with a clock variable with some $Z$ eigenvalue $\zz$, we cannot generate all other eigenvalues by applying $X^d$ to it, we can only generate $r\equiv N/\GCD$ of them. 

At first glance this seems to lead to a macroscopic degeneracy: all the eigenvalues $\zz$ can be divided into $\GCD$ classes, and we are free to choose one member of each class for each site, leading to a total degeneracy of $\GCD^L$. Hamiltonians with such macroscopic degeneracy are poorly defined since their ground state is extremely sensitive to perturbations. 

This is in fact not a problem. Let us first consider the case where $c$ and $d$ are mutually prime. We can apply terms from the Hamiltonian $r$ times to the ground state and see that the following must be true:
\begin{equation}
(\zz_{i-1}^{*})^{rc} (\zz_{i+1})^{rc} \alpha[\zz_1,..., \zz_{L}]=\alpha[\zz_1,..., \zz_{L}].
\label{degen_0}
\end{equation}
For $c$ and $\GCD$ mutually prime, this condition is satisfied if 
\begin{equation}
\zz_{i-1}=\zz_{i+1}\omega^{\GCD\times({\rm integer})}.
\label{degen_condition}
\end{equation} 
Thus we find that $\zz$ of the same species (even or odd) must be equivalent to each other up to a factor of $\omega^{\tilde{d}}$. In other words, after choosing one of $\tilde{d}$ degenerate states for $\zz_0$ and $\zz_1$, all other choices are fixed and the degeneracy is reduced to $\GCD^2$. 

We can also see that
\begin{equation}
(Z_{i-1}^{\dagger c}X_i^d Z_{i+1}^c)^{rm}=(Z_{i-1}^\dagger Z_{i+1})^r,
\end{equation}
where $m$ is some integer satisfying $mc-nd=1$ (which always exists when $c$ and $\GCD$ are mutually prime). We can use this to show that:
\begin{equation}
\bra{\Psi_0}(Z_i^\dagger Z_j)^r\ket{\Psi_0}=1,
\end{equation}
for all sites $i$ and $j$. Therefore there is long-ranged order in the $Z^r$ operators, and since these do not commute with the $\Theta$ operators the $\znn$ symmetry is broken.
However, there is still some symmetry left. The following operators commute with the Hamiltonian, and have no effect when they act on the degenerate ground states:
\begin{equation}
\tilde{\Theta}_{\rm odd}\equiv \prod_{i=2j+1} X_i^{\GCD}, ~~~~ \tilde{\Theta}_{\rm even}\equiv \prod_{i=2j} X_i^{\GCD},
\label{symmetry2}
\end{equation}
These operators generate the symmetry $\zrr$, so when $\GCD>1$, the $\znn$ symmetry is spontaneously broken down to $\zrr$. Using the above arguments, we find that when $\GCD>1$, each species can be divided into $\GCD$ sectors, which are not connected to each other by an operator in the Hamiltonian and which have the same energy. In total there are therefore $\GCD^2$ ground states.
As in higher dimensions, we have found that binding multiple topological defects to a charge leads to a ground state degeneracy. In this one-dimensional case the ground state degeneracy comes not from intrinsic topological order but from spontaneous symmetry breaking. 

When the symmetry is broken down to $\zrr$, there are predicted to be a maximum of $r$ different topological phases,\cite{WenScience} and all of these phases can be realized through different choices of $c$. To see this, we take a similar approach to that above. We insert into the $\tilde{\Theta}_{\rm odd}$ expression identities of the form $Z^{cs}_{2j}(Z^\dagger_{2j})^{cs}$, where $s$ is an integer such that $sd=\GCD$ (mod $r$). It can be easily shown $s$ exists and is mutually prime with $r$. In this case
\begin{equation}
\tilde{\fl}_{\rm odd}=X_1^{\GCD}Z_2^{cs}, ~~ \tilde{\fl}_{\rm even}=(Z_1^\dagger)^{cs}, ~~ \gamma=\omega^{\GCD cs}\equiv \tilde{\omega}^{cs}.
\end{equation}
Defining $\tilde\omega=e^{i2\pi /r}$, we see that $r$ different values of $\gamma$ are possible and all can be realized by different choices of $c$. This implies the existence of $r$ distinct topological phases. Note that there are $N$ values of $c$ but only $r$ topological phases, this is because different $c$'s which are related by adding multiple factors of $r$ realize the same $\gamma$. 
On an open chain, we find a degeneracy of $\GCD^2$ due to the spontaneous symmetry breaking, and a further degeneracy of $(r/\tilde{c}^{\prime})^2$, [where $\tilde{c}^\prime\equiv \gcd(c,r)$] due to the topological properties giving rise to the edge states.

When $c$ and $d$ in the Hamiltonian of Eq.~(\ref{ham1}) are not mutually prime, the argument in Eqs.~(\ref{degen_0})-(\ref{degen_condition}) no longer holds, and we may have macroscopic degeneracy. 
In some cases we can fix this, since, as shown above, models with $c$ different by a multiple of $r$ realize the same $\gamma$. Therefore if a given $c\in[0,r-1]$ is not mutually prime with $d$ we may be able to realize the same $\gamma$ with a different Hamiltonian that has $c$ shifted by $r$ to be mutually prime with $d$.
This still does not allows us to realize phases where $c$, $d$ and $r$ all share a common factor. In this case we need to add an extra term to the Hamiltonian to remove the macroscopic degeneracy:
\begin{equation}
\delta H^{\prime}=-J\sum_i [(Z_{i-1}^\dagger Z_{i+1})^r + h.c.].
\end{equation}
Such a term commutes with the original Hamiltonian, and does not change anything about the ground states when $c$ and $d$ are mutually prime. When $c$ and $d$ are mutually prime, its only effect is to reduce the degeneracy to $\GCD^2$ and make the problem well defined. Therefore we can assume such terms have been added and we do not need to worry about the macroscopic degeneracy. 

\section{Discussion and Conclusion}
One can of course imagine perturbing the Hamiltonian in Eq.~(\ref{ham1}), for instance by adding terms such as $X_i$ or $Z_{i-1}^\dagger Z_{i+1}$. Since the model is one-dimensional we can study the effects of such terms using density matrix renormalization group (DMRG), and in addition the model can also be accessed using sign-free Monte Carlo simulations. We find that the phases described above are stable to such perturbations. The phases can be identified by measuring appropriate string-order parameters,\cite{Quella86} or by computing the projective symmetry group from the entanglement spectrum.\cite{PollmannTurner} 

In summary, we have constructed a class of one-dimensional models with $\znn$ symmetry. These models realize $N$ different topological phases, which respect the symmetry. In addition, when $N$ is not prime the symmetry can be broken down to $\zrr$, where $r$ is a divisor of $N$, and in this case $r$ topologically distinct phases can be realized. Like other models of topological phases, these models work by binding topological defects (domain walls) to charges and proliferating the resulting bound states. The ease of studying our models, which are exactly solvable and can also be easily studied numerically, may make them useful tools for exploring ideas related to the interactions between symmetry and topology. One possible extension would be to study the critical properties of the transitions between these topological phases.

\acknowledgments
We would like to thank V. Quito for useful discussions. 
This research is supported by the National Science Foundation through grant DMR-1206096, and by the Caltech Institute of Quantum Information and Matter, an NSF Physics Frontiers Center with support of the Gordon and Betty Moore Foundation.

\bibliography{1DIsing}
\end{document}